\documentclass[twocolumn,showpacs,preprintnumbers,aps,prb,floatfix,amsmath,amssymb,superscriptaddress]{revtex4-1}
\usepackage{graphicx}
\usepackage[bf,SL,BF]{subfigure}

\begin{document}

%\title{Dissipative dynamics in coupled quantum dots:
%dynamical localization and robust states}
\title{Excitonic entanglement of protected states in quantum dot molecules}
\author{H. S. Borges}
\affiliation{Instituto de F\'{i}sica, Universidade Federal de Uberl\^{a}ndia, 38400-902, Uberl\^{a}ndia-MG, Brazil}
\affiliation{Departamento de F\'{i}sica, Universidade Federal de S\~ao Carlos, 13565-905, S\~ao Carlos-SP, Brazil}
\author{L. Sanz}
\author{A. M. Alcalde}
\affiliation{Instituto de F\'{i}sica, Universidade Federal de Uberl\^{a}ndia, 38400-902, Uberl\^{a}ndia-MG, Brazil}

%\affiliation{Instituto de F\'{i}sica, Universidade Federal de Uberl\^{a}ndia, 38400-902, Uberl\^{a}ndia-MG, Brazil}

\begin{abstract}  
The entanglement of an optically generated electron-hole pair in artificial quantum dot molecules is calculated considering the effects of decoherence by interaction with environment. 
Since the system evolves into a mixed states and due to the complexity of energy level structure, we use the negativity as entanglement quantifier, which is well defined in $d \otimes d^\prime$ composite vector spaces. 
By a numerical analysis of the non-unitary dynamics of the exciton states, we establish the feasibility of producing protected entangled superpositions by an appropriate tuning of bias electric field, $F$. 
A stationary state with a high value of negativity (high degree of entanglement) is obtained by fine tuning of $F$ close to a resonant condition between indirect excitons.
%Stationary behavior of negativity and populations is obtained for a precise adjust of $F$ close to resonance between indirect excitons.  
We also found that when the optical excitation is set approximately equal to the electron tunneling coupling, $\Omega/T_e \sim 1$, the entanglement reaches a maximum value. 
In front of the experimental feasibility of the specific condition mentioned before, our proposal becomes an useful strategy to find robust entangled states in condensed matter systems. 
%This condition can be implemented experimentally and our results can be a very useful strategy to finding robust and entangled states in solid state systems.

\end{abstract}
 \maketitle

\section{Introduction}
\label{sec:intro}
Semiconductor quantum dot molecules (QDMs) driven by coherent pulses have been extensively suggested as promising candidates for physical implementation of solid state quantum 
information processing~\cite{Gammon03,Economou12}. The possibility of selective control of the electronic occupation and a 
controllable energy spectrum are the major characteristics which allow QDMs to be viable for the realization of universal 
quantum computation~\cite{Gammon06}. The flexibility of quantum dots (QDs) as quantum information systems has been proven by successful implementation of controllable operations 
on charge~\cite{Gorman05} and optical qubits~\cite{Biolatti00,Mathew11}. 

In addition to a feasible qubit physical implementation, quantum entanglement is a fundamental nonlocal resource for quantum computation and communication. 
Although there are several theoretical approaches proposing methods for direct measurement of entanglement~\cite{Horodecki02,Santos06}, its experimental quantification remains elusive.
Optically induced entanglement of excitons has been performed in single QDs~\cite{Chen00} and QDMs~\cite{Bayer01}, where interaction between particles and interdot tunneling are the key mechanisms for the formation of entangled states. 
%The experimental quantification of entanglement remains elusive, primarily because it is not an observable which allows for the direct measurement.  
Several theoretical calculations have proposed different strategies to obtain exciton states in QDMs with a high degree of entanglement~\cite{Nazir04, Paspalakis04, Chu06}. Also, the electron-hole entanglement can be efficiently tuned and optimized through the proper choice of interdot separation, QD asymmetry and by the action of an electric field applied in the growth direction~\cite{Bester04,Bester05}. 
%In ideal conditions, quantum entanglement it is not affected by decoherence during the quantum information processing and its calculation can be easily performed by means of von Neuman entropy~\cite{Chu06}. 

In ideal conditions, the QDM undergoes unitary evolution, the quantum entanglement is not affected by decoherence and its calculation can be easily performed by using Von Neumann entropy~\cite{Chu06,Bester04}. 
However, a system is unavoidable coupled with the environment which leads to degradation of quantum coherence. 
In these conditions, the system evolution is non-unitary and the decoherence effects cause deterioration of the entanglement and will be detrimental for production, manipulation and detection of entangled states. 
For QDMs the main decoherence channels are the radiative decay of excitons and exciton pure dephasing which is important even at low temperatures~\cite{Bardot05}. 
%While in closed system the Von Neumann entropy can be used as entanglement measurement~\cite{Chu06}, 
The definition of a computable quantifier of entanglement for general mixed states in open quantum systems has been a challenge for the last decades~\cite{Plenio07}.
%In an open quantum system, the system is in a mixed state, and this fact difficult the definition of a computable measurement of entanglement for generic mixed states~\cite{Plenio07}. 
Among the diverse bipartite entanglement measurements, entanglement of formation~\cite{Valle11} and concurrence~\cite{Wootters97} are well-defined and extensively used to evaluate the entanglement of mixed bipartite in $2 \otimes 2$ systems. 
%For general vector spaces $d \otimes d^\prime$ systems, the generalization of these measurements for general mixed states remains a challenging problem. 
%some quantifiers the
%Entanglement of Formation $E(\rho)$~\cite{Valle11}. Based on the expression demonstrated by Wooters, for two-qubit states this entanglement can be obtained by a computable quantify called concurrence $C(\rho)$~\cite{Wootters97}. 
An alternative measure for mixed states is negativity, $\mathcal{N}(\rho)$, first proposed by Vidal and Werner~\cite{Vidal02}, which overcomes the limitations 
of other measurements and allows to calculate the degree of entanglement for general $d \otimes d^\prime$ composite vector spaces~\cite{Lee03}. 
It is important to mention that, in the context of QDMs, electron-hole entanglement was previously investigated using the von Neumann entropy ignoring the essential effects of decoherence~\cite{Bester04,Bester05,Chu06}. 
%The negativity is based on the partial transposition criterion for separability and quantify how much a given composed state $\rho$ fails to violate this criterion. 
%This measure has the advantage to be easily evaluated and to quantify entanglement not only between two-qubits but also, between subsystems of $l$-levels ($l>2$). 

In this paper, we investigate the entanglement degree of electron-hole pairs created in QDMs by the incidence of coherent radiation considering spontaneous exciton decay and pure dephasing as main decoherence sources. The degree of entanglement in the asymptotic regime is evaluated through the negativity $\mathcal{N}(\rho)$ as a function of controllable physical parameters, exploring the conditions which maximize the entanglement degree. 
Our results show that the system evolves to asymptotic states induced by dissipative mechanisms which are superpositions of indirect exciton states. For experimentally accessible conditions, such states have a long lifetimes~\cite{Borges10} and high degree of entanglement. 

\section{Description of the System}
\label{sec:theory}
We consider a QDM composed by two asymmetric QDs vertically aligned and separated by a barrier of width $d$. The electron-hole occupation is controlled by the interplay of tunneling coupling, optical excitation and gate potentials. 
Due to structural asymmetry of QDs, the energy levels of each carrier become resonant for specific values of an external electric field $F$ applied along the growth direction of the QDM. 
A careful growth engineering along with the natural QDM asymmetry allows the design of samples with selective tunneling of electrons or holes~\cite{Bracker06}.
 
%A careful sample design and an appropriate choice of $F$ govern the electron or hole tunneling in QDMs~\cite{Bracker06}. 
In order to investigate the entanglement between electron and hole, we model our system using a composite particle position basis 
$|X^{e_{\mathrm{T}},h_{\mathrm{T}}}_{e_{\mathrm{B}},h_{\mathrm{B}}}\rangle = 
\vert^{e_\mathrm{T}}_{e_\mathrm{B}}\rangle \otimes
\vert^{h_\mathrm{T}}_{h_\mathrm{B}}\rangle$, 
where $e_{\mathrm{T}(\mathrm{B})},h_{\mathrm{T},(\mathrm{B)}}$ represent the occupation number of electrons and holes in each level in the top (T) or bottom (B) QD, respectively~\cite{Rolon10,Bayer01,Korkusinski2002610}. 
The QDM is driven by a low-intensity continuous wave laser, such that only the ground-state exciton can be formed. The occupation of the electron or the hole in each QD should be 0 or 1, where the value 0 (or 1) represents the absence (presence) of the carrier in the QD. For instance, the ket $\vert X^{10}_{01}\rangle$ represents an indirect exciton state, with one electron occupying the top QD and one hole in the bottom QD. 
Thus, the QD position index encodes the information of a specific quantum state and the complete basis set of the composite system $H=\mathcal{H}_e \otimes \mathcal{H}_h$ is comprised by 16-states. 
Considering only the optical active transitions, tunneling of electrons and holes and assuming that the QDM is initially uncharged, the QDM system is described using the composite basis 
$\{\vert X^{00}_{00}\rangle$, 
$\vert X^{00}_{11}\rangle$,
$\vert X^{10}_{01}\rangle$,
$\vert X^{11}_{00}\rangle$,
$\vert X^{01}_{10}\rangle$,
$\vert X^{11}_{11}\rangle\}$.
Under electric-dipole and rotating-wave approximations and after removing the time dependence, the resulting Hamiltonian is given by %(taking the energy of ground states as energy origin, $\varepsilon_0=0$ and $\hbar=1$):
\begin{eqnarray}
\label{eq:1}
H=\left(
\begin{array}{cccccc}0&\Omega&0&\Omega&0&0\\
\Omega&\delta_{_{11}^{00}}&T_e&V_f&T_h&\Omega\\
0&T_e&\delta_{_{01}^{10}}-\Delta_F&T_h&0&0\\
\Omega&V_f&T_h&\delta_{_{00}^{11}}&T_e&\Omega\\
0&T_h&0&T_e&\delta_{_{10}^{01}}+\Delta_F&0\\
0&\Omega&0&\Omega&0&\delta_{_{11}^{11}}+V_{XX}
\end{array}
\right),
\end{eqnarray} 
where $\delta_{^{e_\mathrm{T},h_\mathrm{T}}_{e_\mathrm{B},h_\mathrm{B}}}$ is the detuning of the incident laser and the exciton states,  
%between the energy of each state and the energy of the incident laser ($\hbar\omega_L$) and
$\Delta_F = eFd$ is the Stark energy shift on the indirect excitons~\cite{Zrenner02}, being $d$ the barrier thickness between the QDs. 
$T_{e(h)}$ describes the single-particle interdot tunneling for electrons (holes), $V_f$ is the interdot coupling between direct excitons via the F\"{o}ster mechanism, and $V_{XX}$ accounts for direct Coulomb binding energy between two excitons, one located on each dot~\cite{Lovett02,Rolon07}. 
% (induces an exciton transfer from one to another dot) 
The optical coupling is given by the parameter    
%by We assume the presence of a single laser with Rabi frequency 
$\Omega$, which depends on the laser intensity and oscillator strength of allowed optical transitions. 
%electronic structure of the semiconductor, coupling the ground state to direct exciton states in each dot. The Fig.\ref{fig1} represents schematically the excitons energy levels and the couplings of our model.

For numerical calculations we use the exciton bare energies and coupling parameters given in Refs.~\cite{Rolon10,Bracker06} for an InAs/GaAs QDM. 
\begin{figure}[htb]
\centering
\includegraphics[scale=0.6]{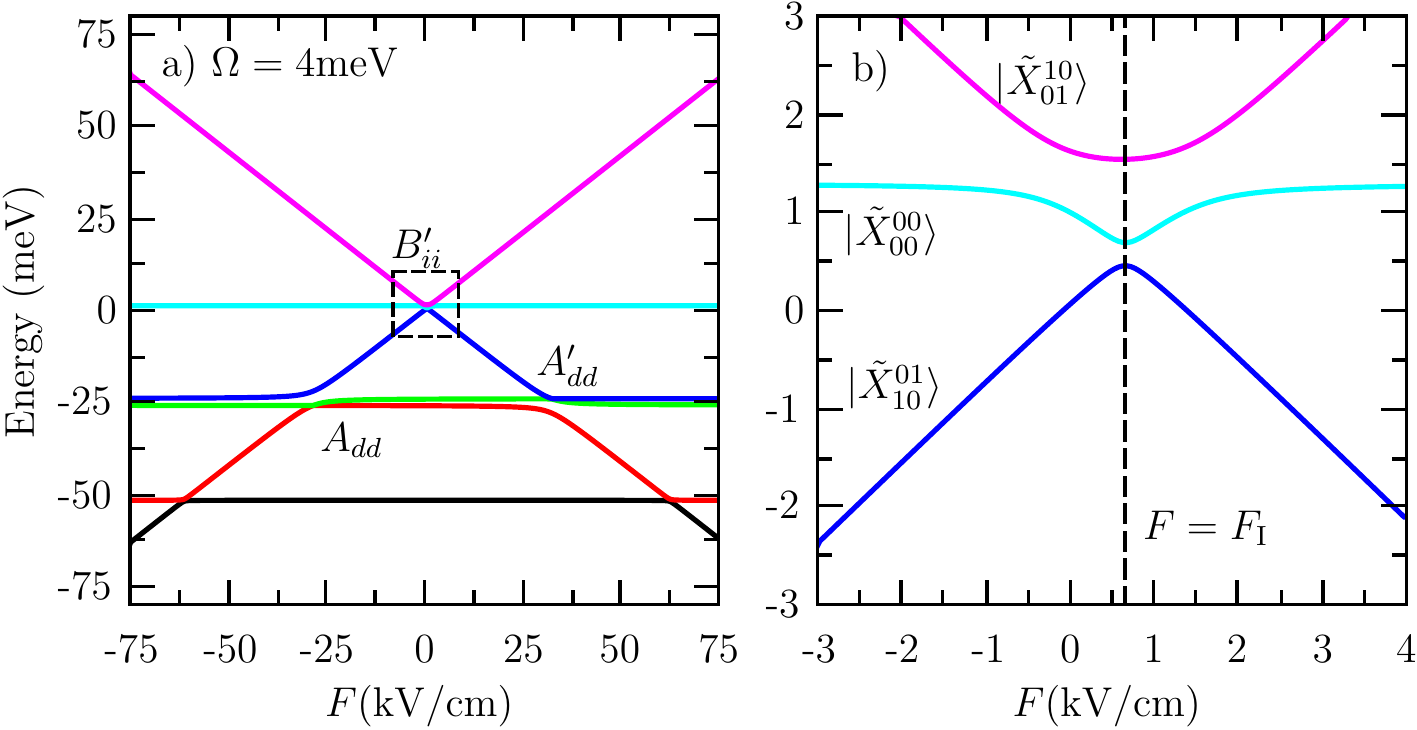}
\caption{(color online) Left panel: Energy eigenvalues of Hamiltonian (\ref{eq:1}) as a function of electric field $F$, 
for $\Omega = 4$meV, $V_f = 0.08$meV, $T_e = 20T_h = 2$meV, $V_{XX} = -5$meV and $d = 8.4$meV. 
Right panel: Zoom of the dashed area in panel a), showing the energy eigenvalues near to $F=F_\mathrm{I}$ which corresponds to the anticrossing condition between the indirect exciton states. 
For simplicity, we label the exciton vacuum $\vert \tilde{X}_{00}^{00}\rangle$ and indirect exciton states 
($\vert \tilde{X}_{10}^{01}\rangle$, $\vert \tilde{X}_{01}^{10}\rangle$) levels according to their dominant character for large $F$. 
}
\label{figure1}
\end{figure}
Figure~\ref{figure1}a) shows the eigenvalues of Hamiltonian (\ref{eq:1}) as a function of electric field $F$, for 
$\Omega = 4$meV and $d=8.4$meV. 
The energy levels consist of direct exciton states, weakly dependent on the electric field $F$, and indirect exciton states, which are strongly dependent with $F$. 
Direct and indirect exciton states are coupled by tunneling and exhibit large anticrossings. 
At well-defined values of $F$, it is important to note the arising of small anticrossings, labeled as $A$ and $B$ in Fig.~\ref{figure1}a). These anticrossings correspond to coupling between exciton states of the same kind. 
For instance, the anticrossings identified as $A_{dd}$ and $A^\prime_{dd}$ are related to the coupling between direct exciton states (intradot excitons). 
From here on, we focus our attention at anticrossing labeled as $B_{ii}$. In Fig.~\ref{figure1}b), we show the detailed structure of anticrossing $B_{ii}$ which involves the two indirect exciton states and the vacuum state. 
The indirect exciton states (interdot excitons) $\vert \tilde{X}_{10}^{01}\rangle$ and $\vert \tilde{X}_{01}^{10}\rangle$ are effectively coupled at field value $F = F_\mathrm{I}$. 
This particular value of the electric field is obtained through of the resonance condition between indirect bare excitons: $\delta^{10}_{01}-\Delta_{F}=\delta^{01}_{10}+\Delta_{F}$.  

Although the coupling mechanism between indirect exciton states cannot be distinguished directly from Hamiltonian~(\ref{eq:1}), this is a result of the combined action of both, electron and hole tunnelings. This can be checked by projecting out the direct exciton states of the total Hamiltonian to obtain the effective coupling between indirect excitons which is found to be proportional to $T_eT_h$. It is also found that the effective coupling between the vacuum state and indirect excitons is nearly proportional to $\Omega T_{e(h)}$. A detailed analysis of the underlying mechanisms behind of indirect excitons coupling will shed light on the optimal construction of robust entangled states.   
%It is important to note that the structural QDM parameters, as QDs separation distance $d$, have direct influence on the numerical values of the electric field where anti-crossings occurs. 
%Using the basis associated to number occupation (0 or 1) for each dot of the molecule, the individual system (electron or hole) has four possible states. For instance, the electron can %be in the bottom (top) dot $|X^{0(1)}_{1(0)}\rangle$ or can be present (absent) in both dots $|X^{1(0)}_{1(0)}\rangle$. However, considering the intra-band and inter-band %transitions within our assumptions and assuming the both quantum dots are uncharged before of the laser be turned on, only the six states of exciton basis of the Hilbert space of the %composite system ($\mathcal{H}_e \otimes \mathcal{H}_h$) are accessible. 
\section{Entanglement in QDM}
%\subsubsection{Closed system}
The anticrossings of type $A$ and $B$ are directly related to the emergence of strong electron-hole entanglement as shown in Refs.~\onlinecite{Chu06,Bester05}. 
In closed quantum systems, this assertion is proven through the calculation of von Neumann entropy,  defined as $S=-\mathrm{Tr}\rho_e \log_2 \rho_e = -\mathrm{Tr}\rho_h \log_2 \rho_h$, where  $\rho_{e,(h)} = \mathrm{Tr}_{h,(e)}(\rho_{eh})$ is the reduced density matrix for electron (hole). 
For a system whose dimension is $d$, the maximum value of entropy is $S_{max} = \log_2(d)$, which corresponds to maximally entangled states. 
In closed quantum systems, the density matrix of the composite electron-hole system, $\rho_{eh}$, is obtained from the von Neumann equation: $i\hbar \frac{\partial}{\partial t} \rho_{eh}= \left[ H,\rho_{eh}\right]$.  
The entropy $S$ calculated as a function of electric field $F$ is composed by narrow peaks located at field values where the anticrossings of type $A$ and $B$ occur~\cite{Chu06}. 

We perform a numerical calculation of entropy $S$ for optical excitation $\Omega = 4$meV at field values corresponding to anticrossings of type $A$ and $B$. At values of $F$ where anticrossing of type $A_{dd}$ occurs, the obtained entropy $S$ is approximately 55\% of $S_{max}$. At $F=F_I$, corresponding to indirect states anticrossing $B_{ii}$, the von Neumann entropy $S$ attains 75\% of its maximum value $S_{max}$. 
This high degree of entanglement observed at anticrossing $B_{ii}$ is interesting for two reasons: i) indirect exciton superpositions with large entanglement in QDM can be engineered at $F = F_\mathrm{I}$ by a suitable choice of hamiltonian parameters, and ii) in previous work~\cite{Borges10}, we proved that at the condition $F = F_\mathrm{I}$, the system evolves to an asymptotic superposition of vacuum and indirect exciton states, which is protected against decoherence. 
At this point, it is important to stress that the controlled production of entangled states as robust quantum superpositions is one of the key ingredient in the design of any quantum information system. 
Whereas the emergence of robust states in QDM is determined by the competitive effects of decoherence and tunneling, it is necessary to determine if the large values of entanglement are preserved under the same decoherence mechanisms. 
Establish the conditions for a controlled generation of entangled states protected against decoherence is the goal of the next section. 

%\subsubsection{Open system}
In order to investigate the dissipative effects on the exciton dynamics and entanglement, we solve the Liouville-Von Neumann-Lindblad equation given by:
\begin{equation}\label{eq:master}
i \hbar \frac{\partial}{\partial t} \rho_{eh}\left(t\right)=
\left[H,\rho_{eh}\left(t\right)\right]+ i \hbar L(\rho_{eh}\left(t\right)).
\end{equation}
Here, the Liouville superoperator, $L(\rho) = L_\mathrm{D}(\rho) + L_\mathrm{I}(\rho)$, which describes dissipation effects due to recombination and pure dephasing mechanisms for direct (D) and indirect excitons (I), can be written as: 
%\begin{eqnarray}\label{eq:liouville}
%L(\rho)^\mathrm{D(I)} &= & 
%\sum_{j} \frac{\Gamma_\mathrm{D(I)}}{2}
%\left(
% 2 \sigma_-^j \rho \sigma_+^j - \sigma_+^j \sigma_-^j \rho - \rho \sigma_+^j \sigma_-^j 
%\right) + \nonumber \\
%& &  
%\sum_{k} \gamma_\mathrm{D(I)}
%\left(
% 2 \sigma_z^j \rho \sigma_z^j - \sigma_z^j \sigma_z^j \rho - \rho \sigma_z^j \sigma_z^j  
%\right),
%\end{eqnarray}
\begin{eqnarray}\label{eq:liouville}
L_\mathrm{D(I)} & = & 
\sum_{j}
\frac{\Gamma^0_\mathrm{D(I)}}{2}
\left(
 2 \vert v \rangle \langle j \vert \rho \vert j \rangle \langle v \vert - \vert j \rangle \langle j \vert \rho - \rho \vert j \rangle \langle j \vert
\right) +  \nonumber \\
& &  
\sum_{j} 
\gamma_\mathrm{D(I)}
\left(
 2 \vert j \rangle \langle j \vert \rho \vert j \rangle \langle j \vert - \vert j \rangle \langle j \vert \rho - \rho \vert j \rangle \langle j \vert   
\right), 
\end{eqnarray}
where the index $j$ runs over the direct exciton states for $L_\mathrm{D}(\rho)$ and indirect exciton states for $L_\mathrm{I}(\rho)$. 
$\Gamma^0_\mathrm{D(I)}$ is the decoherence rate associated to spontaneous decay from the optical excited state $\vert j\rangle$ 
to exciton vacuum $\vert v\rangle \equiv \vert X^{00}_{00}\rangle$, while $\gamma$ describes pure dephasing in each excitonic level $\vert j \rangle$.  
We use the effective rates given by $\Gamma_\mathrm{D(I)} = \Gamma^0_\mathrm{D(I)}/2 + \gamma_\mathrm{D(I)}$ and $\Gamma_\mathrm{I} = 10^{-3} \Gamma_\mathrm{D}$, where $\Gamma_\mathrm{D}=10\mu$eV~\cite{Borri03,Bardot05}. 

Nowadays, the determination of a general entanglement measurement for open quantum systems is the subject of an intense theoretical debate, particularly for Hilbert spaces of dimension $d \otimes d$ $(d > 2)$. Several entanglement quantifiers for mixed states have been proposed, whose algebraic implementation involves different degrees of complexity. In order to determine the degree of entanglement for our multilevel QDM considering decoherence effects we 
use the measure known as negativity, $\mathcal{N}(\rho)$, which is well defined for arbitrary dimension $d$ and whose calculation is obtained from the numerical solution of (\ref{eq:master}). The negativity for a general composite $A-B$ system of dimension $d \otimes d$ is defined 
as~\cite{Vidal02,Lee03}
\begin{equation}
\mathcal{N}(\rho_{AB}) = \frac{\left \| \rho_{AB}^{T_{A}} \right \|_1 - 1}{d-1}, 
\label{negativity}
\end{equation}
where $\rho_{AB}^{T_{A}}$ is the partial transpose of a state $\rho_{AB}$ with respect to subsystem $A$ and $\left \| \cdot \right \|_1$ represents the trace norm. The negativity, as defined in~(\ref{negativity}), is normalized so that the maximum value of negativity is $\mathcal{N}_{max} = 1$. 

\begin{figure}[htb]
	\centering
	\includegraphics[scale=0.8]{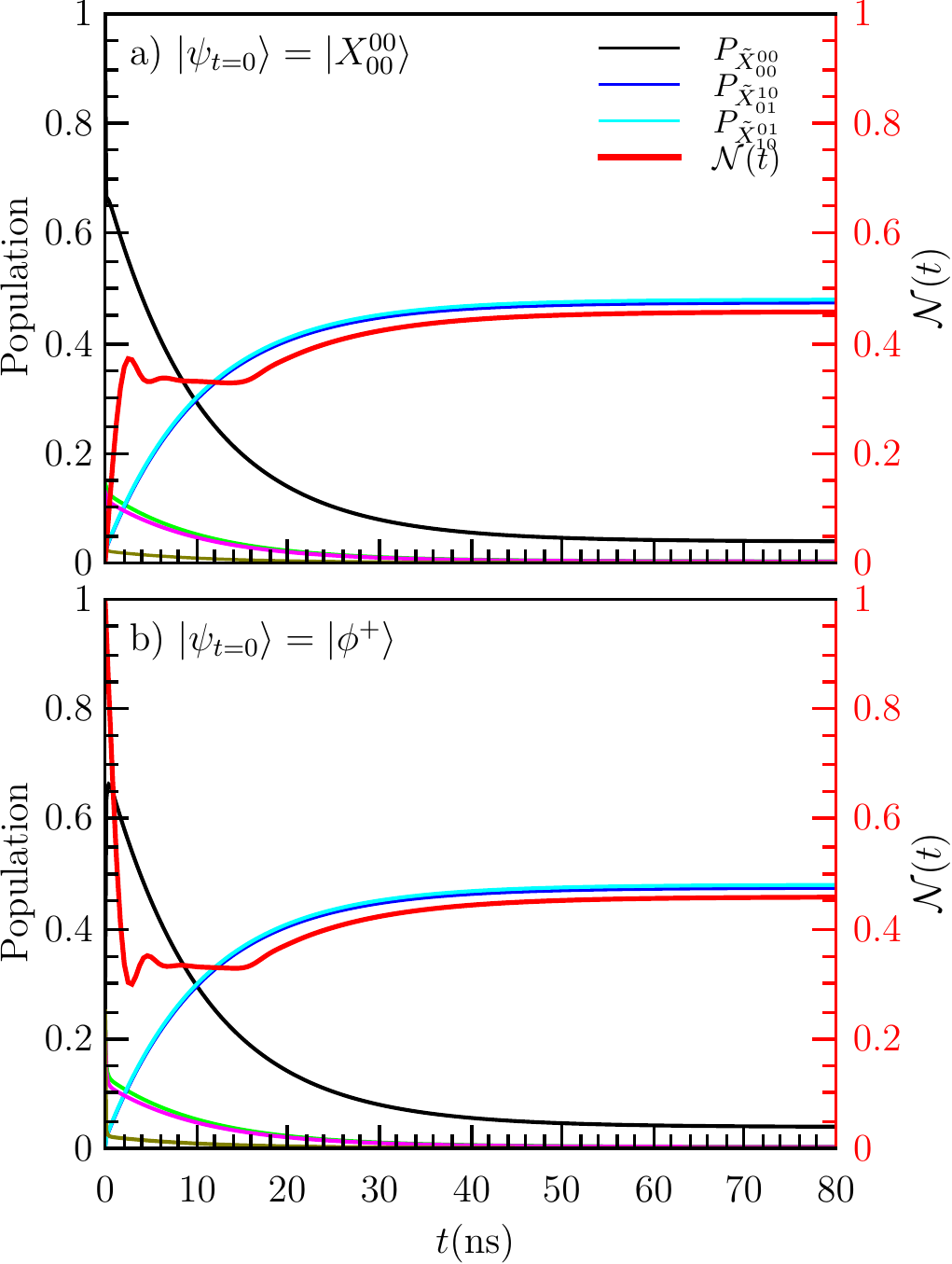}
	\caption{(color online). Time evolution of populations $P_j(t)$ for two different initial states (left axis) and negativity $\mathcal{N}(t)$ (right axis) in red solid line. The system evolves to a superposition of indirect excitons with a small contribution of exciton vacuum. The optical excitation is $\Omega=10$meV, the remaining of parameters are the same used in Fig.~\ref{figure1}. The indirect exciton resonance condition is obtained at $F_\mathrm{I} = 0.64$kV/cm.  
	}
	\label{figure2}
\end{figure}
The evolution of the system to an asymptotic excitonic superposition at $F = F_\mathrm{I}$ is verified by solving numerically the equation system~(\ref{eq:master}).  
Figure~\ref{figure2} shows the populations, $P_j=\rho_{jj}(t)$ of the $j$ state, for two different initial states $\psi(t=0)$. The upper panel corresponds to the case when the system is initially prepared in exciton vacuum state $\psi(t=0) = \vert X^{00}_{00}\rangle$ and the lower panel is for one of the maximally entangled states of our system $\psi(t=0) = \vert \phi^+ \rangle = \frac{1}{2} (\vert X^{00}_{00} \rangle + \vert X^{10}_{01} \rangle + \vert X^{01}_{10} \rangle + \vert X^{11}_{11} \rangle)$~\cite{Wilde13}. In both cases we use the coupling $\Omega = 10$meV and $F_\mathrm{I} = 0.64$kV/cm. On a short-time scale, the populations exhibit fast oscillations compatible with their corresponding initial states. For sufficiently long times, $t\gg 1/\Gamma_\mathrm{D}$, each population reaches the same stationary value independent of the choice of the initial state. 
For the particular set of parameters used in this calculation, the system evolves to an asymptotic state $\rho_{st}$ formed mainly by a superposition of indirect excitons with $P_{\tilde{X}^{10}_{01}} \sim  P_{\tilde{X}^{01}_{10}} \sim 0.48$, and a small contribution of the vacuum. 
We can compare the steady state $\rho_{st}$ with the pure superposition $\vert \psi_p \rangle = \frac{1}{\sqrt{2}} \left(\vert X^{01}_{10} \rangle + \vert X^{10}_{01} \rangle \right)$ through the fidelity $\mathcal{F} = \mathrm{Tr}\left( \rho_{st}, \rho_p \right)$, which gives $F\sim 96~\%$. 
%Comparing the obtained steady state $\rho_{st}$ with the pure superposition $\vert \psi_p \rangle = \frac{1}{\sqrt{2}} \left(\vert X^{01}_{10} \rangle + \vert X^{10}_{01} \rangle \right)$ through the fidelity $\mathcal{F} = \mathrm{Tr}\left( \rho_{st}, \rho_p \right)$, where $\rho_p = \vert \psi_p \rangle \langle \psi_p \vert$. Evaluation of fidelity $\mathcal{F}$ yield a value of 96~\%. 
%Further calculations of fidelity with respect to the eigenstates of Hamiltonian~(\ref{eq:1}) for different rates $\Omega / T_{e,h}$ and $F=F_\mathrm{I}$ show that the system evolves asymptotically with high fidelity into an eigenstate of the system. 
On the right axis of Fig.~\ref{figure2} we show the negativity (red solid line) as a function of time. For asymptotic time scales, the negativity reaches a constant value ($\mathcal{N} \sim 0.45$) independently of the choice of the initial state. The robustness of entanglement is a remarkable result because it demonstrates the feasibility of producing protected entangled states by a simple tuning of the electric field at $F=F_\mathrm{I}$. 
\begin{figure}[htb]
	\centering
	\includegraphics[scale=0.8]{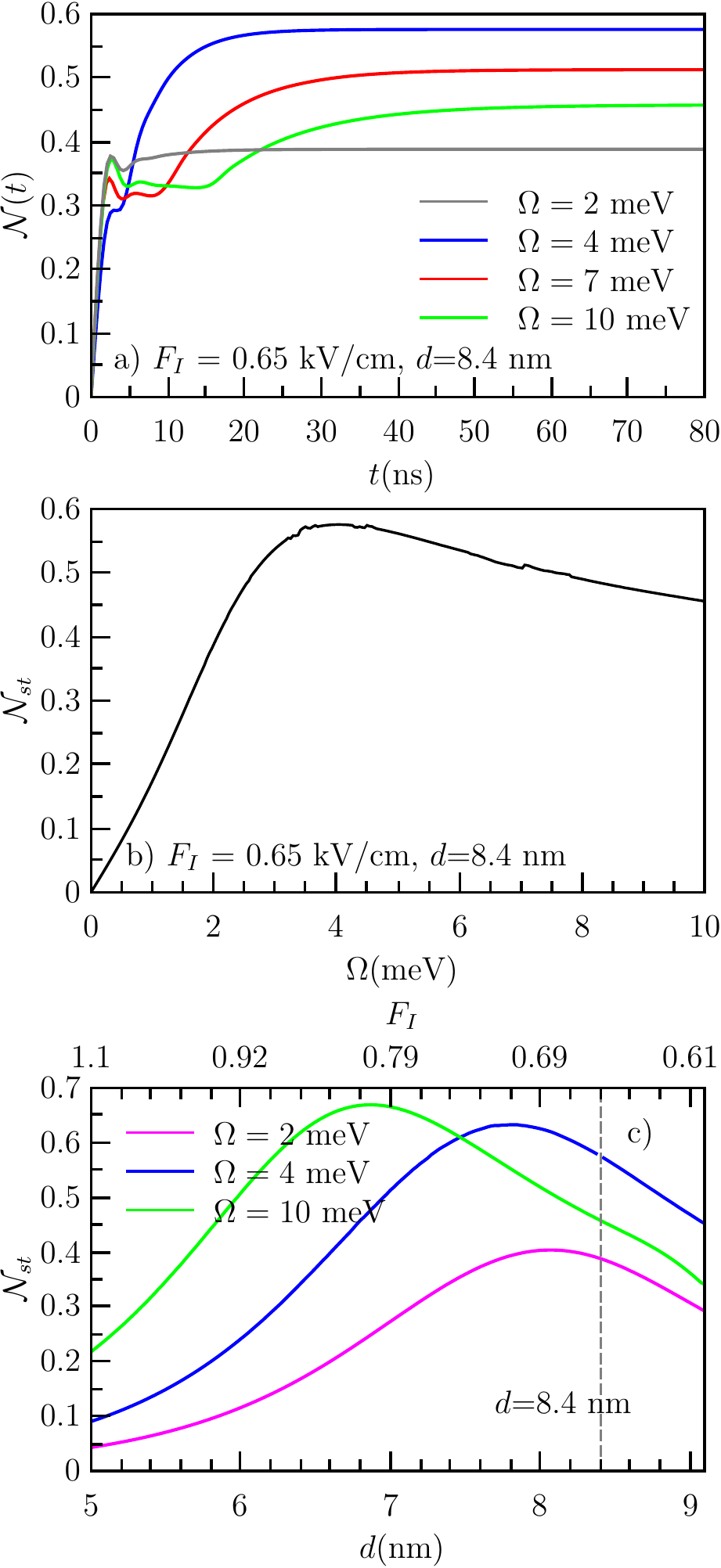}
	\caption{(color online). a) Negativity as a function of time for different optical excitations $\Omega$. For long-time dynamics the negativity reaches a stationary value $\mathcal{N}_{st}$. b) Asymptotic negativity, $\mathcal{N}_{st}$,  as a function of optical coupling parameter. c) $\mathcal{N}_{st}$ as a function of QD separation $d$ for different values of $\Omega$. In this case, the value of $F_I$ and tunneling parameters are calculated for each value of $d$. For comparison with above results, the vertical dashed line indicates the distance $d=8.4$nm. 
	}
	\label{figure3}
\end{figure}

We continue examining the entangled asymptotic states in Fig.~\ref{figure3}a), where we show the time evolution of negativity for several choices of optical coupling. Again, for sufficiently long times, when the system reaches its stationary regime, the negativity evolves to a steady value, $\mathcal{N}_{st}$. Some interesting facts should be pointed out: i) likewise the exciton population, the stationary behavior of negativity is a direct consequence of the actions of decoherence processes. 
%This sort of entanglement assisted by decoherence takes place at time scales proportional to the ratio $ T/\Gamma^0_\mathrm{D}$. 
ii) the time required to reach the stationary value $\mathcal{N}_{st}$ is a linear increasing function of $\Omega$ with an approximate slope $\sim T/\Gamma^0_\mathrm{D}$, iii) conversely, $\mathcal{N}_{st}$ does not exhibit a linear dependence with $\Omega$. 
To achieve a better comprehension of the relation between the stationary negativity $\mathcal{N}_{st}$ and optical excitation, in Fig.~\ref{figure3}b) we show $\mathcal{N}_{st}$ as a function of $\Omega$ for $F = F_\mathrm{I}$ and considering the same parameters used above. For small optical excitations, the asymptotic negativity $\mathcal{N}_{st}$ increases almost linearly with $\Omega$ up to a maximum value. Then, the degree of entanglement decays as $\Omega$ continues to increase. For the particular set of parameters used in the calculation, the maximum value of $\mathcal{N}_{st}$ is obtained at $\Omega = \Omega_m \approx 4$meV. 

It is also interesting to show the relation between the degree of entanglement and the asymptotic states of the system. Analyzing the density matrix elements, we verify that the asymptotic state has a general form given by $\vert \psi_a\rangle \sim c_0 \vert X^{00}_{00} \rangle + c_1 \vert X^{10}_{01} \rangle + c_2 \vert X^{01}_{10}\rangle$ with small contributions of other states. Depending on the value of $\Omega$, we can distinguish three different behaviors: for $\Omega <\Omega_m$, the component $c_0$ associated with vacuum is dominant in the formation of the asymptotic state. If $\Omega > \Omega_m$ the components $c_1$ and $c_2$ associated with the indirect exciton states are dominant, being approximately equal $c_1 \sim c_2$. For the particular case of $\Omega \sim \Omega_m$, the vacuum and the two indirect excitons contribute in approximately equal weight to the formation of the asymptotic state. 
Thus, the long-lived state formed by the superposition of indirect excitons does not necessarily lead maximum entanglement. In fact, when $\Omega$ is chosen such that the vacuum is populated in the same proportion that indirect states, we obtain maximum entanglement.

The population of indirect exciton states is controlled by the interplay of decoherence rate and the ratio $\Omega/T_e$. 
An analysis based on the effective Hamiltonian of the reduced system formed by vacuum and two indirect excitons shows that when $\Omega/T_e \sim 1$ it is possible to populate the vacuum and indirect states in the same proportion and therefore we can obtain optimal entangled states. 
To verify this assertion, in Fig.~\ref{figure3}c) is shown $\mathcal{N}_{st}$ as a function of the interdot distance, $d$,  which is directly related with $T_e$ through $T_e \sim e^{-d}$ (Ref. \onlinecite{tunnel}). For all considered optical excitations $\Omega$, we noted that negativity reaches a maximum value at well-defined values of the interdot separation $d$. 
These interdot distances that optimize the negativity decreases as $\Omega$ increases. Note that, the resonant field $F_\mathrm{I}$ as well as the tunneling rate $T_e$  are calculated for each value of $d$. Thus, for $\Omega =2$meV, the maximum value of negativity is obtained when $d=8.1$nm, the corresponding electron tunneling rate is $T_e = 2.5$meV, resulting in $\Omega/T_e = 0.8$. Similar results are obtained for $\Omega =4$meV ($\Omega/T_e = 1.1$), and $\Omega = 10$meV ($\Omega/T_e = 1.09$). 
Experimentally, the condition $\Omega/T_e \sim 1$ can be implemented using optical spectroscopic techniques. 
To achieve a stationary state with the maximum value of $\mathcal{N}_{st}$ using a QDM sample with interdot separation $d$, one has to determine the tunneling coupling and then to adjust the optical excitation according with $\Omega=T_e$.
%For a particular QDM sample characterized by the interdot separation $d$, the tunneling coupling must be determined and the optical excitation adjusted according to $\Omega = T_e$. 
The experimental determination of tunneling couplings can be done from photoluminescence measurements of the anticrossing energy gap $\Delta$ between direct and indirect exciton states~\cite{Bracker06,Doty09}. 
Following a simple two-level approximation, the tunneling coupling $T$ for both electrons and holes can be obtained by $2T = \Delta$. 
We point out that for all cases studied, the hole tunneling coupling $T_h$ does not affect significantly the dynamics of the system.
%Thus, we provide a reasonably efficient method for the optical tailoring of protected states with an appreciable degree of entanglement.

\begin{figure}[htb]
\centering
\includegraphics[scale=0.6]{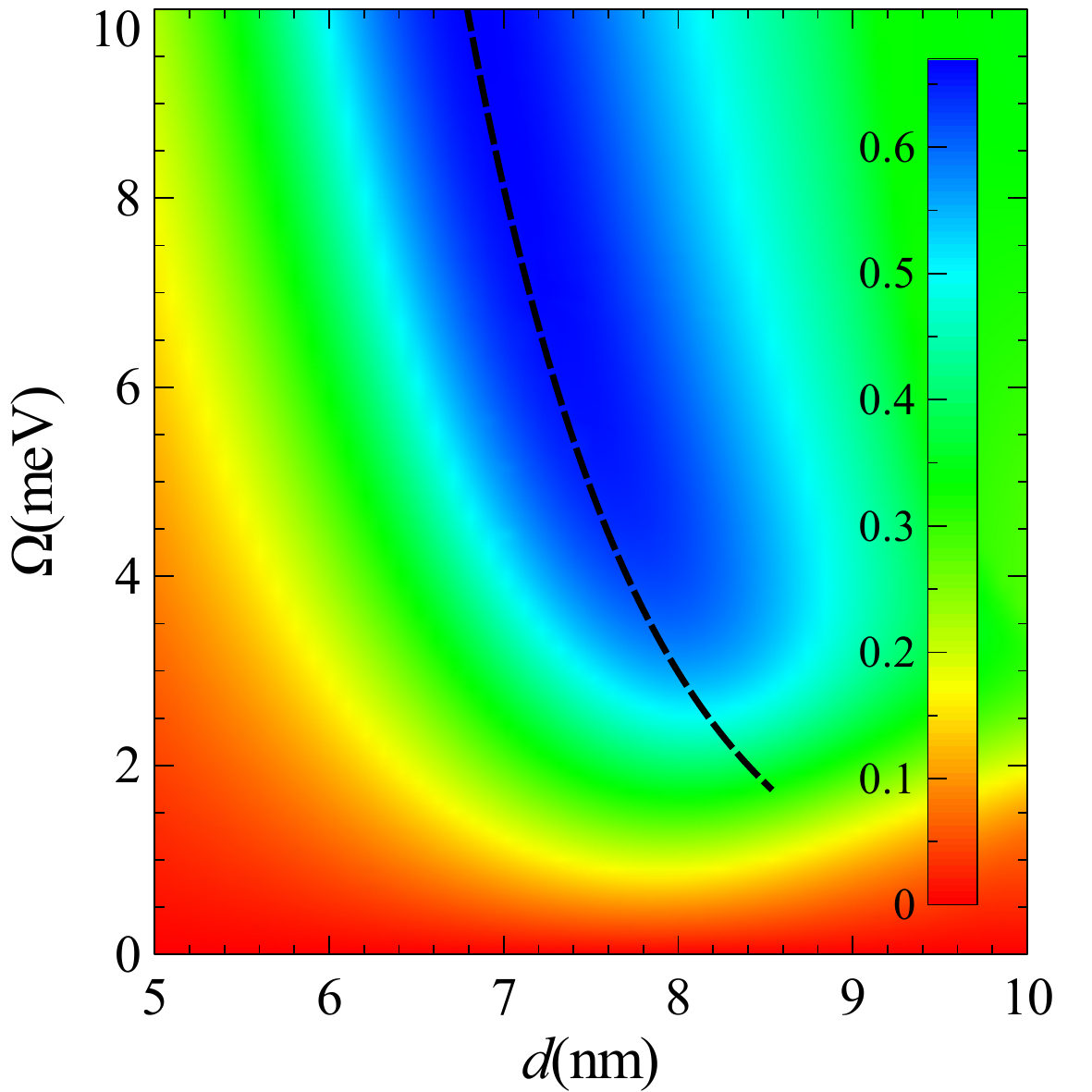}
\caption{(color online). Stationary negativity as a function of interdot separation $d$ and $\Omega$.  Optimal entangled superpositions of type $\vert \psi_a\rangle \sim c_0 \vert X^{00}_{00} \rangle + c_1 \vert X^{10}_{01} \rangle + c_2 \vert X^{01}_{10}\rangle$ with $c_0 \sim c_1 \sim c_2$ are obtained following the condition $T_e \sim \Omega \sim e^{-d}$ displayed by the black dashed line. 
}
\label{figure4}
\end{figure}

We summarize our findings in Fig.~\ref{figure4}, where we show the stationary negativity $\mathcal{N}_{st}$ as a function of interdot separation $d$ and optical coupling $\Omega$.
The values of $\mathcal{N}_{st}$ are calculated keeping the condition $F = F_\mathrm{I}$ for each interdot distance $d$. As discussed above, the maximum values of $\mathcal{N}_{st}$ can be obtained following the condition $\Omega \sim T_e \sim e^{-d}$. This means that the $\Omega$ values that maximize the negativity must have the same dependence on $d$ that tunneling $T_e$. We confirm this assertion by plotting in black dashed line the same empirical relationship used to adjust the tunneling $T_e = A_e + B_e e^{-d/D_e}$~\cite{tunnel}. We can see that for typical excitations range, the $\Omega$ values that maximize the negativity (dark blue region) follow approximately the same dependence on $d$ as the tunneling $T_e$.
%As is showed in Refs.~\cite{Chu06, Bester04, Bester05}, the anti-crossings observed for different electric field values in the energy excitonic levels spectrum, is a clear signature of this coupling between the levels. 
%Through of the suitable tuning of electric field applied along of the growth direction, we can control the quantum coupling between the different exciton states, and this feature in turn influence the degree of the entanglement existing between electron and hole. 
%. For our investigation, we consider realistic parameters for InAs self-assembled QDM under coherent laser excitation and a gate electric field.  
%The parameters associated to the molecule structure and excitons energies was based in the measurements performed and investigated in the Refs.~\cite{Krenner05,Chu06}.
%Differently other papers we investigate the decoherence effects on the entanglement excitonic and optimized their     

%\begin{figure}[htb]
%\centering
%\includegraphics[scale=0.3]{fig2.jpg}
%\caption{(a) Energy exciton spectrum  and (b) entropy of entanglement for each eigenstate, as function of electric field.}\label{fig2}
%\end{figure}

\section{Conclusions}
\label{sec:Conclusions}
%We establish the conditions that allow obtain a stationary behavior of negativity and found a parameter regime that maximizes the negativity.
%We also provide a experimental strategy, which allows to obtain robust entangled states in artificial molecules based on quantum dots.
From the analysis of the open system, we found a regime where it is possible to obtain superpositions of states with a high degree of entanglement.
We use this parameter regime as a guide to investigate whether the entanglement remains robust to the effects of decoherence. 
In the open system, we found that the dynamics of the QDM converges to a superposition with a high degree of entanglement and simultaneously protected against decoherence.
From our results, it is important to recall that the high degree of entanglement is sustained by the interplay of vacuum and decoherence. 
The role of the vacuum state becomes clearer if we consider that one of the maximally entangled states of the system does not have the simple form of a Bell state, but rather $\phi^+ = \frac{1}{2} \left(\vert X^{00}_{00}\rangle + \vert X^{10}_{01} \rangle + \vert X^{01}_{10} \rangle + \vert X^{11}_{11} \rangle\right)$. Thus, the $B_{ii}$ anticrossing, involves the largest number of necessary states for the formation of a superposition with a high degree of entanglement.
We establish the conditions that allow obtaining a stationary behavior of negativity and found a parameter regime that maximizes the negativity providing a viable experimental strategy to obtain robust entangled states in artificial molecules based on quantum dots.

\begin{acknowledgments}
The authors gratefully acknowledge financial support from Brazilian Agencies CAPES, CNPq, and FAPEMIG. H.S.B. acknowledges support from FAPESP (2014/12740-1). This work was performed as part of the Brazilian National Institute of Science and Technology for Quantum Information (INCT-IQ). 
\end{acknowledgments}
%\bibliographystyle{unsrt}
%\bibliographystyle{apsrev}
%\bibliography{borgesnegativity}

\begin{thebibliography}{10}
\bibitem{tunnel}
	In general, the tunneling coupling and the interdot distance are related by
	$T_{e,h} = A_{e,h} + B_{e,h}\exp(-d/D_{e,h})$. In order to fit the
	experimental results given in Ref.~\cite{Bracker06} for electrons (holes) we
	use: $A_e$ = -0.037meV ($A_h$=-0.253meV), $B_e$ = 8.0meV ($B_h$ = 1.048meV)
	and $D_e$ = 5.848nm ($D_h$ = 7.446nm).
\end{thebibliography}

\end{document}